\input{aipcheck}

\documentclass[
  ,draft            
  ]
  {aipproc}

\layoutstyle{6x9}

\newcommand{\pasp}{PASP}
\newcommand{\apj}{ApJ}
\newcommand{\aap}{A\&A}

\newcommand{\apjs}{ApJS}
\newcommand{\mnras}{MNRAS}
\newcommand{\apjl}{ApJL}
\newcommand{\nat}{Nature}
\newcommand{\aj}{AJ}
\newcommand{\araa}{ARAA}

\begin{document}

\title{Models of Accretion Disks}

\classification{90. 95. 95.30.Lz 95.85.Mt 95.85.Nv 97.10.Gz 97.60.Lf 98.62.Mw 98.54.Aj 98.54.Cm \texttt{http://www.aip..org/pacs/index.html}}

\author{Aneta Siemiginowska}{
  address={Harvard-Smithsonian Center for Astrophysics, 60 Garden St., Cambridge,MA 02138 USA}
}


\begin{abstract}

An accretion flow onto a supermassive black hole is the primary
process powering quasars. However, a geometry of this flow is not well
constrained.  Both global MHD simulations and observations suggest
that there are several emission components present in the nucleus: an
accretion disk, hot plasma (corona or sphere) with electrons
scattering the optical and UV photons, and an outflow (wind/jet). The
relative location and size of these emission components, as well as
their ``interplay'' affect the emerging quasar spectrum.  I review
briefly standard accretion disk models and the recent progress, point
out discrepancies between the predicted and observed spectra and
discuss some issues in fitting these models to the broad-band spectral
energy distribution of quasars. I present examples of models fitted
simultaneously to the optical-UV-X-ray data and possible constraints
on the parameters.

\end{abstract}

\maketitle


\section{Idea of an Accretion Disk}

Images posted on many Astronomical Web Sites\footnote{for example
\url{http://chandra.harvard.edu/photo/} \\
\url{http://hubblesite.org/newscenter/archive/releases/}} present artist's
views of the central engine of quasars, active galactic nuclei (AGN), and
binary systems (XRB). They usually contain a black dot, representing a black
hole, surrounded by a disk-like structure ``showing'' a circular
motion of matter flowing towards the black hole potential. There is
also a perpendicular to the disk funnel representing a jet.
Such artists views of the central engines are not exactly the images
astronomers can see in the real observations. Usually, even the
highest resolution images show simply a diffuse or in most cases a
point-like emission. The incredibly high luminosity ( $> 10^{12}
L_{\odot}$ for quasars) emitted by such an unresolved point source is
the main indication of the powerful physical processes involved in
generating this energy.

The idea of an accretion disk is almost 50 years old. After the
discovery of quasars as luminous radio sources the obvious question to
ask was how they are powered? Salpeter (1964) \cite{salpeter1964}
suggested that an accretion onto a massive black hole and a release of
the gravitational energy associated with that process can provide this
power.  Lynden-Bell (1969) \cite{lynden-bell1969} gave an idea of ``a
matter swirling down the Schwarzschield mouth'' and presented a
concept of an accretion process (and a picture!) that might be at
work. In the following years, several papers describing physics of an
accretion onto a massive black hole have been published (for example
\cite{lynden-bell1971,pringle1972,prp1973})
It is remarkable to read these papers today and
reflect on the way they have defined one of the most important
processes in astrophysics in our times.

In 1973 Shakura \& Sunyaev \cite{ss1973} considered the physics of
accretion in great detail including a discussion of energy
dissipation, an angular momentum transport by both viscosity and
outflow, an origin of turbulence as magnetic or convective. They
presented a parameterization of viscosity ($\alpha$-viscosity) in
terms of efficiency of the angular momentum transfer.  The nature of
the viscosity has been the key problem in the accretion disk theory,
while this convenient parameterization allowed for progress in both
analytical modeling and applications to the observations.  Twenty
years later, Balbus \& Hawley (1991) \cite{balbus1991} show that the
magnetic turbulence associated with MRI (magnetorotational
instability) is the best candidate for the viscosity. \footnote{An
excellent review of the angular momentum transport and MRI is given by
Balbus (2003) \cite{balbus03}}. Now, we understand that accretion
flows are magnetohydrodynamical and the magnetic fields play crucial
role in the driving the accretion. However, the main difficulty in the
analytical description of such disks means that we need to develop
simulations to understand the accretion process. While the huge
progress in numerical simulations has been made over the last decade
there is still no full 3D global accretion flow model available. The
accretion phenomenon is still not well understood and there are many
remaining questions still waiting for an answer.

Here, I consider issues related to the application of accretion disk
models to quasar spectra and constraints on the accretion flow
geometry based on the spectral fitting of quasar data.

\section{Observational Tests}

Testing accretion models in quasars has been challenging. We cannot
directly image the accretion flow because the central 1~pc region of a
galaxy remains unresolved.  Therefore in understanding the accretion
process we mainly depend on observations of the spectral energy
distribution (SED) and variability, and indirectly on large scale
signatures of the quasar activity such as jets and outflows. Modeling
the observed spectra from the accretion flow however is non-trivial
because the disk broadband optical-UV and X-ray emission has to be
taken into account, while the interstellar absorption affects the bulk
of this emission and hides the peak of the emission, the key feature
for the disk temperature determination in quasars and AGN. Thus the
both ends of the spectral energy distribution needs to be taken into
account to constrain the model parameters in a meaningful way.

\subsection{Geometry of Accretion Flow}

The geometry of accretion flow is not well understood and there is no
critical observational constraints available at this time.  In
particular, we do not understand whether the flow becomes spherical,
how and where the corona forms above the disk, whether the corona
covers the disk at all times, etc.  Global MHD simulations (with MRI
but no radiation) indicate that several distinct regions form within
an accretion flow onto a supermassive black hole
\cite{haw2002,proga2003}: a Keplerian turbulent disk enveloped by a
hot and unconstrained corona, a hot inner torus, a magnetically
confined unbound jet and a centrifugal tunnel. However, these global
simulations do not account for radiation.  Note that to calculate the
spectrum emerging from an accretion flow one needs to take into
account emission from each region as well as irradiation and
self-illumination of these regions. Thus the global models needs to be
developed. Some progress is being made and results of local MHD
simulations, which include radiation effects have been recently
reported (see \cite{hirose2006,blaes2006,krolik2006}).

Elvis (2000) \cite{elvis2000} considered observational information to
derive an empirical model of a quasar structure. He describes physical
location of the regions associated with different spectral components
(in particular emission lines). The centrifugal funnel and the outflow
are the key features in this model.

\begin{figure}[h]
  \includegraphics[height=.28\textheight]{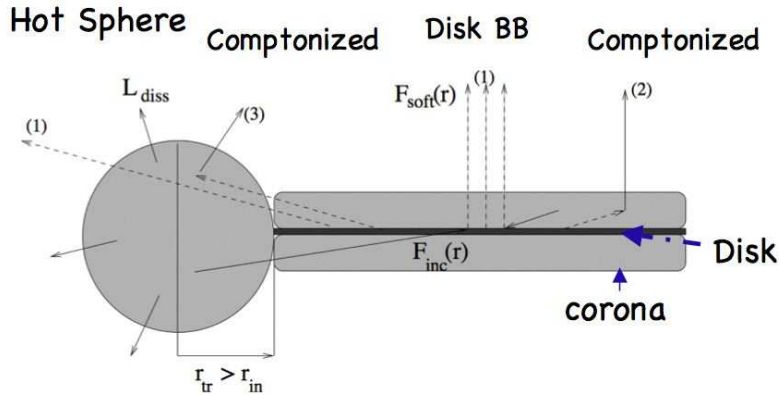}
\caption{Two types of accretion flow geometry \cite{sobol2004a}: 
(1) The hot sphere surrounds
the central black hole and the accretion disk covered by a tenuous
uniform corona extends outside the sphere. The cold disk photons are
inverse Compton scattered in the hot sphere and the corona. Model
parameters: size of the sphere, energy dissipation and temperature of
the corona, mass and accretion rate.}
\label{model1}
\end{figure}

\begin{figure}[h]
  \includegraphics[height=.28\textheight]{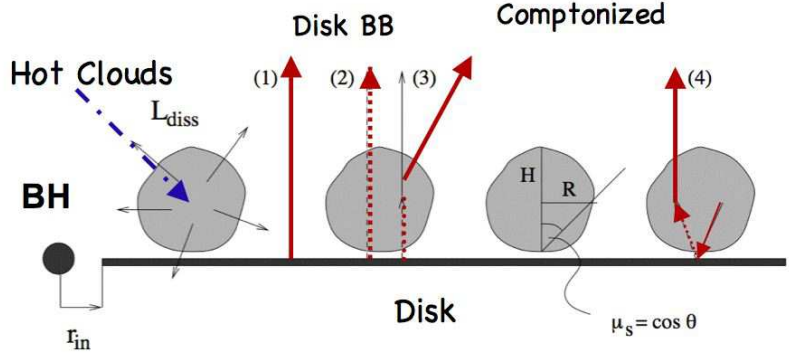}
\caption{Two types of accretion flow geometry \cite{sobol2004b}: 
(2) Geometrically think disk
extends down to the innermost stable orbit of a central black
hole. The patchy corona is located above the disk with hot clouds
representing flare regions. The cold photons are Compton scattered in
the clouds. Model parameters: mass, accretion rate, clouds geometry
and velocity, fraction of energy dissipated in the corona. }
\label{model2}
\end{figure}

\subsection{Emission from an Accretion Disk.}

In 1978 Shields \cite{shields78} suggested that the excess emission (a
``bump'') observed above a powerlaw extrapolation between the radio
and optical continuum in 3C~273 might be a result of a thermal
emission from the disk around a supermassive black hole. Malkan \&
Sargent (1982) \cite{malkan1982} tested this idea with better data and
considered several emission components including the disk blackbody
emission. The standard disk blackbody emission model is conveniently
independent on the energy dissipation process, thus it is independent
on the viscosity. The disk effective temperature is proportional to
the radius, so $T(r) \sim T_0\,r^{-3/4}$ with the normalization scaling
with black hole mass and accretion rate (see
\cite{frank02} for detailed equations). 
This spectral peak is located in X-rays for stellar mass black hole
systems with $\sim 10 M_{\odot}$, and it is shifted to UV for a
supermassive black hole in AGN. The emitted flux increases with the
mass.  A direct application of this model to the spectra of many
quasars and local active galaxies has been possible with general
constraints on the mass and accretion rates.

However, as shown by Czerny \& Elvis \cite{czerny1987} and Hardt \&
Maraschi \cite{hard1991} the opacities of the disk atmosphere and
effects of a tenuous corona above the disk have to be taken into
account when calculating the quasar spectra. Thompson and Compton
scattering of the UV photons on the electrons in the disk atmosphere
or corona, as well as the bound-free opacities significantly modify
the emission. The ``modified'' blackbody disk models show a change in
the spectral shape in the optical-UV and an additional contribution to
the emission in the soft X-rays (
e.g. \cite{ross1992,matt1993,liz1995,fiore1995,fiore1998,chiang2002}). In
addition inclination and GR effects shift the spectra towards higher
frequencies \cite{sun89,laor89,sie90,sie95}. Recently more
sophisticated models have been developed to treat properly the disk
structure and effects of the corona \cite{roz99,rozanska99}, radiative
transfer in accretion disk atmosphere
\cite{hubeny2000,blaes01,hubeny2001}, and irradiation of the
atmosphere
\cite{madej2000,madej2000b} in the disk around a supermassive
black hole.

\section{Spectral Components}

The observed broad-band spectra must originate in different parts of
the accretion flow. This is because the optical-UV spectra usually
associated with a thermal disk emission require much lower
temperatures than the thermal X-ray emission. Thus, at least a range
of temperatures between $10^4-10^6$K is required for the entire
optical-to-X-ray range emission. However, the temperature of the
standard thermal disk is not high enough to contribute significantly
to a typical energy range between 0.1~keV and $\sim 40$~keV (rest
frame). On the other hand, the Compton effect becomes important at
these energies and the X-ray emission is usually associated with the
inverse Compton scattering of low frequency photons (optical and UV)
on electrons in a hot medium. The amplification of the effective boost
of the photon energy towards X-rays requires electron temperatures
between a few to hundreds keV. Depending on the nature of the hot
plasma, the electrons can have thermal or non-thermal
distributions. The main idea is that we should have a source of cold
optical-UV photons and a hot plasma to create the observed spectra.
The geometry of these two medium has not been constrained so far.

Recently Sobolewska et al \cite{sobol2004a,sobol2004b} attempted to
constrain a parameter space of accretion models for quasars
\cite{bech2003}. They considered two types of geometries: (1) an
ionized sphere surrounded by an accretion disk covered by a hot
tenuous corona (Fig.~\ref{model1}); (2) an accretion disk extending to
the last stable orbit covered by a patchy corona with magnetic flares
(Fig.~\ref{model2}). Given the observed spectra of high redshift
quasars ($z>4$) the patchy corona model was more favorable, although
one cannot rule-out either possibility. A consistent modeling of large
quasar samples with accretion models is needed.

Historically, many studies of large samples used a parameterized form
of $\alpha_{ox}$ \cite{tan1979} which describes the relative optical
and X-ray emission.  $\alpha_{ox}$ is defined as a ratio of optical to
UV luminosity thus this parameterization avoids the difficulty related
to modeling different types of data obtained by optical and X-ray
telescopes.  This parameterization is convenient for studies of
evolution and redshift dependence or general properties of the sample
(e.g. \cite{bech2003,strateva2005,kelly2006,steffen2006}). It also
shows the significance of the disk emission in respect to the hot
Comptonizing medium. However, $\alpha_{ox}$ alone is not good enough
for analysis of properties of accretion flow. This is clearly visible
from analysis of the theoretical spectral shape in the optical-UV
band. To calculate $\alpha_{ox}$ one usually needs to extrapolate the
flux at 2500\AA\ or 1450\AA\ and assume the optical-UV slope. However,
the slope in the optical-UV is very sensitive to the properties of the
accretion flow.  Observational results of large quasar samples rarely
give an observed optical-UV spectral slope. The quasars SED compiled
by Elvis et al
\cite{elvis94} for only about 100 quasars has been widely used in many
studies of AGN samples. The new SED for a large SDSS sample gives a
wider scatter around the mean of the optical-UV slope described by
Richards et al \cite{richards2006}.

\begin{figure}[h]
  \includegraphics[height=.3\textheight]{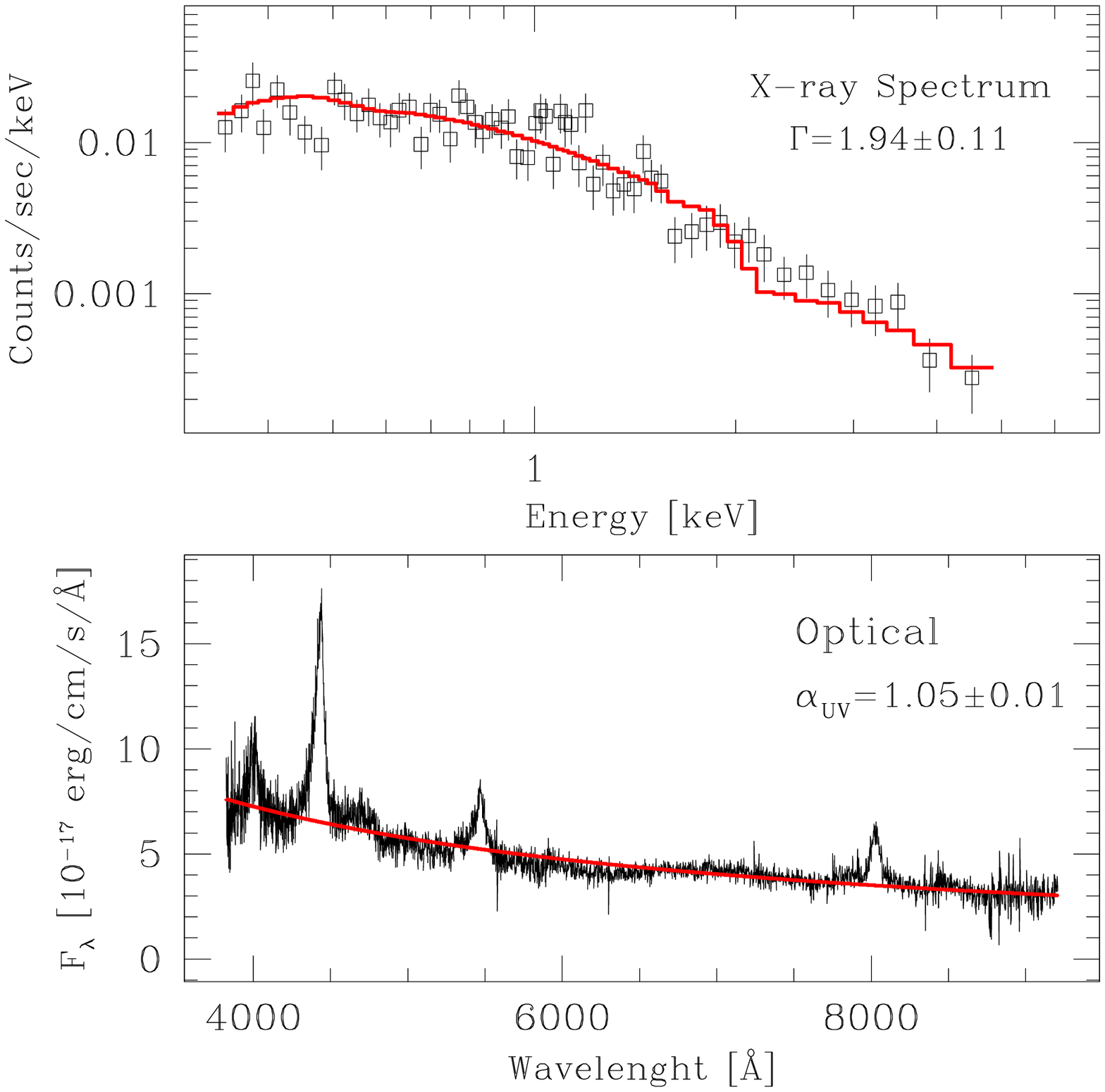}
  \includegraphics[height=.25\textheight]{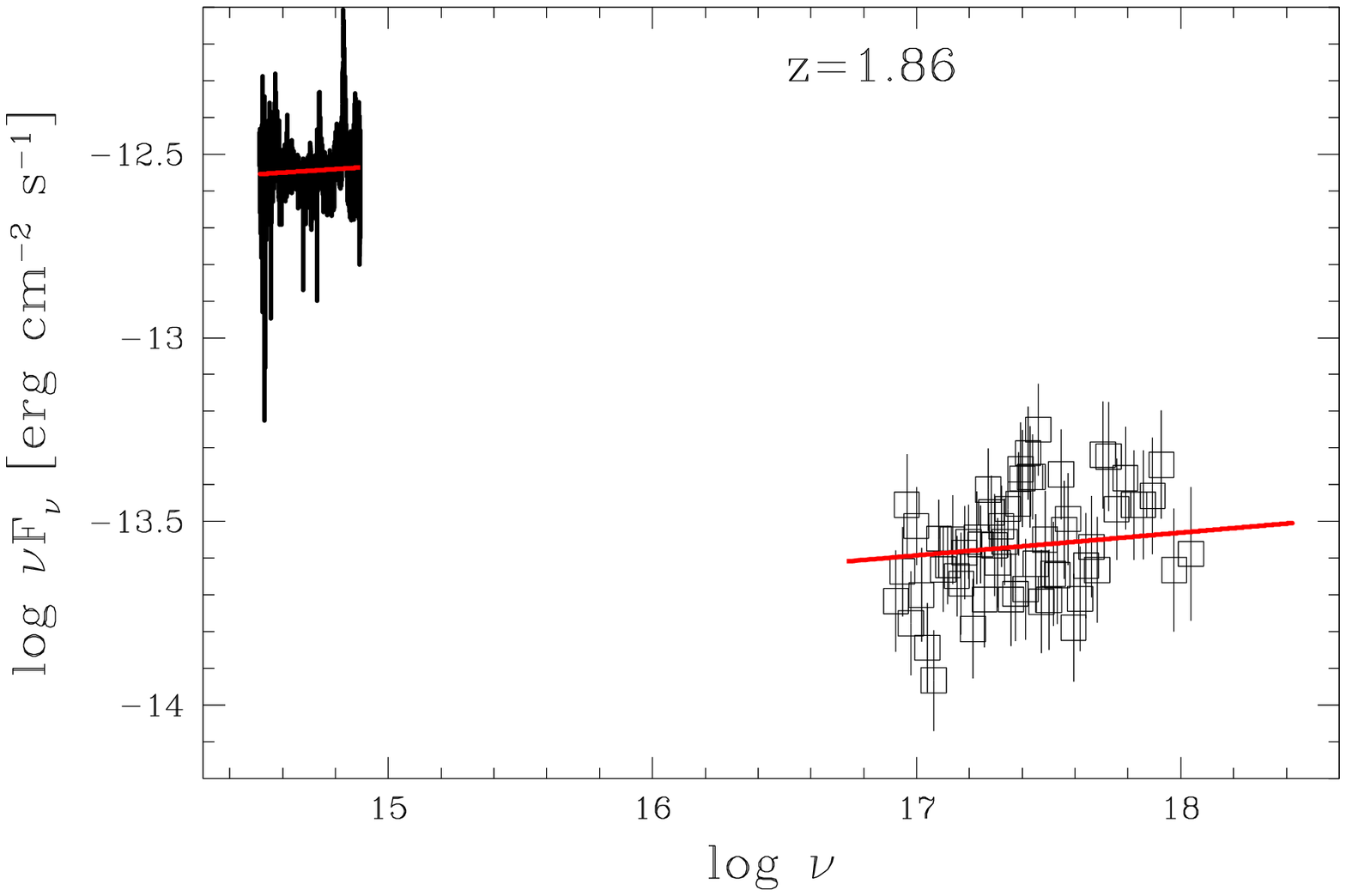}
\caption{Fitting the optical and X-ray spectra. Left: The optical and X-ray
spectra of z=1.86 quasars are fit by different power law models. Top
panel shows the typical X-ray data marked by squares with the error
bars over-plotted with a power law model indicated by a solid
line. The X-ray photon index $\Gamma=1.94^\pm0.11$ and optical slope
$\alpha_{UV} = 1.05\pm 0.01$. Right: the same data on $Log \nu$ vs $
Log \nu F_{\nu}$ plane with the power law models.}
\label{powlaw}
\end{figure}

\begin{figure}[h]
  \includegraphics[height=.25\textheight]{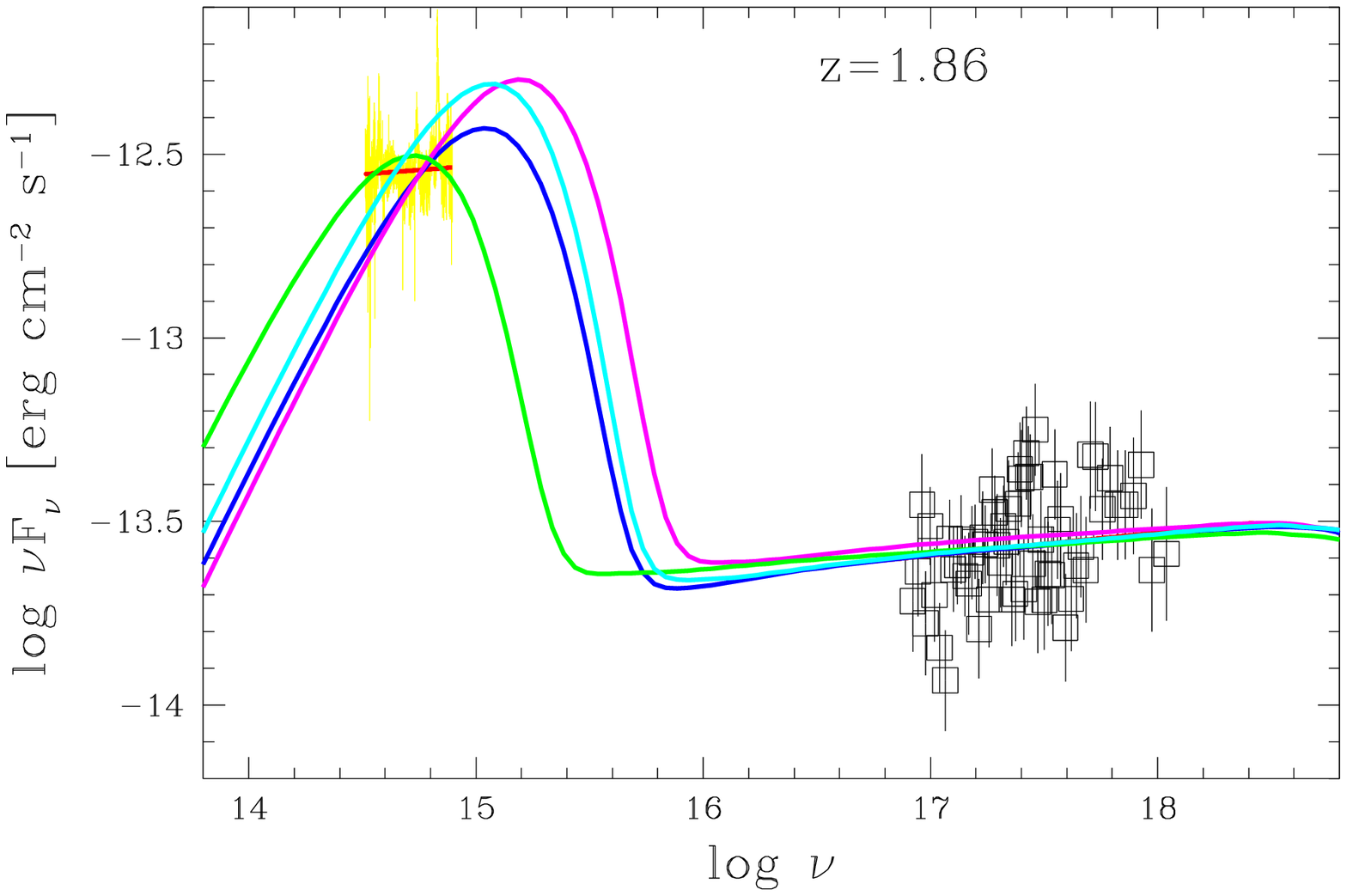}
  \includegraphics[height=.25\textheight]{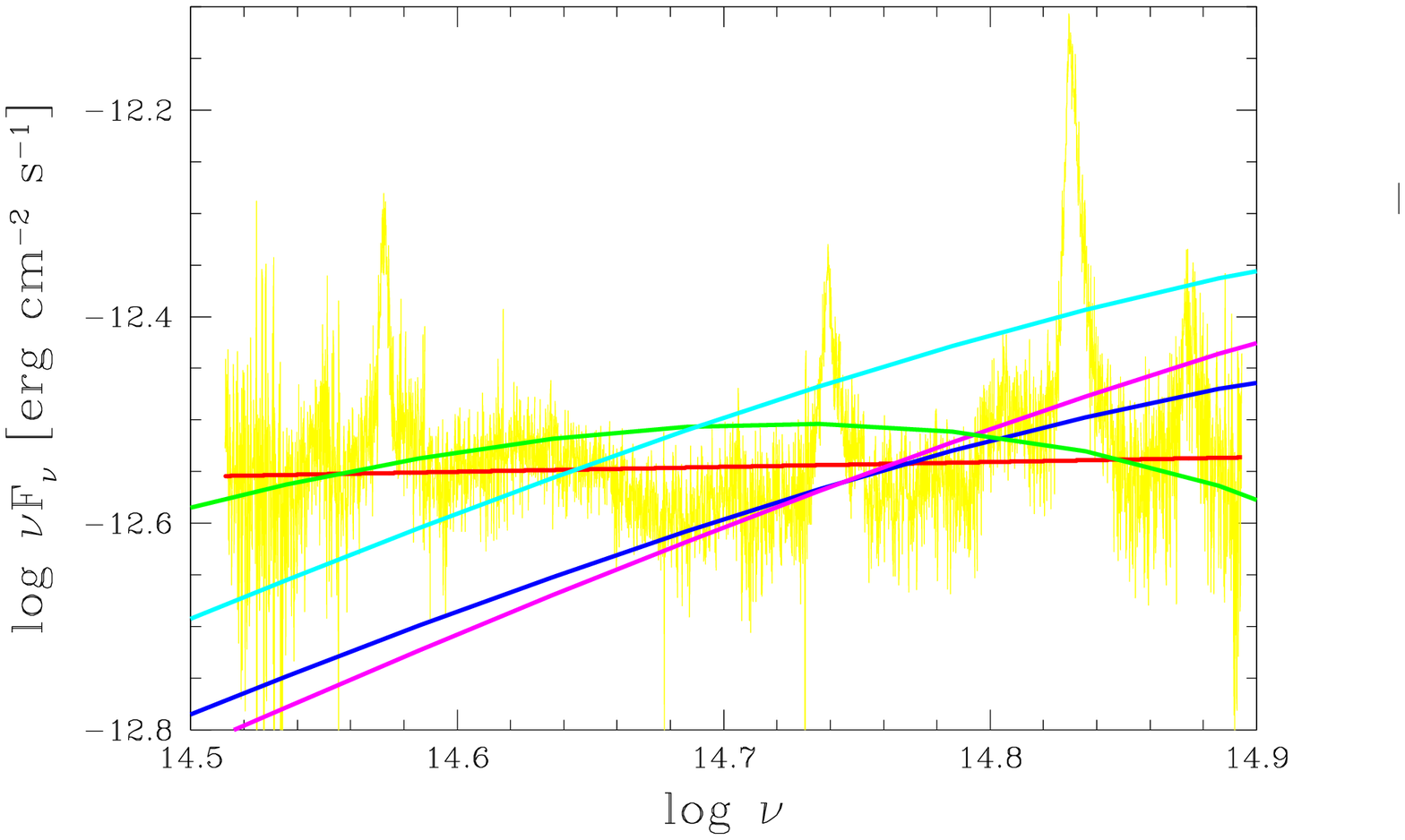}
  \caption{Fitting accretion disk models of Sobolewska et al to the
  broad-band spectral energy distribution of a z=1.86 quasar.  {\bf
  Left:} The broad-band optical-X-ray data plotted with a solid yellow
  line in the optical (emission lines are visible above the continuum)
  and squares with 1~$\sigma$ error bars in the X-rays. The solid
  colored curves indicate different model parameter fits to the
  data. Notice that the models have the same $\alpha_{ox}$ and the
  X-ray slope, but differ a lot in the optical-UV slopes. {\bf Right:}
  The zoom into the optical band to show the differences between the
  models. Model parameters: $M_{BH} = 10^9 M_{\odot} \, $with $\, \dot
  M = 0.4 \dot M_{Edd} \,$\, and $0.5 \dot M_{Edd};$ \, $M_{BH} =
  4\times 10^9 M_{\odot}$ and $\dot M = 0.09 \dot M_{Edd};$ \, $M_{BH}
  = 5.7 \times 10^8 M_{\odot}$ and $\dot M = 1 \dot M_{Edd}$. The red
  line shows the power law fit to the data. }
\label{fig-models}
\end{figure}

\section{Fitting Disk Models to the Data}

For meaningful tests of accretion models one then needs to consider
the full optical-UV-X-ray spectral range.  A lack of a general fitting
package for simultaneous X-ray and optical spectral analysis is a
major hurdle in the wide application of the accretion models to the
data. XSPEC \cite{xspec1996} contains several models, however they are
mainly designed for XRB systems
\cite{li2005,davis2005,davis2006}. Sherpa
\cite{freeman2001} in CIAO software allows for easy use of both
optical and X-ray data, but it does not have a model applicable to AGN
and quasars. Here, we use Sherpa to demonstrate a possible approach to
model fitting and argue that good accretion models should be developed
for general use by the community, since this is the best way to learn
about accretion.

Figure~\ref{powlaw} shows the high redshift quasar spectrum fit in
Sherpa with two separate power law models in the optical-UV and
X-rays. The two ends of the spectrum have been treated independently
giving a parameterized description of the spectrum that is not
related to accretion models.  Figure~\ref{fig-models} shows an example
of the quasar SED fit with a range of accretion flow models. Note that
many models fit quite well the X-ray data, while the optical-UV slope
is missed by most of the models. In this case the model included a
modified blackbody component to account for the accretion disk
emission in the optical-UV and the Comptonization in the patchy corona
contributing to the X-rays. Clearly theoretical improvement in
optical-UV model is needed. One can argue that the optical-UV part of
the spectral model is critical to understanding the geometry of
accretion flows as well as the accretion physics.

Koratkar and Blaes \cite{koratkar1999} considered application limits
of the standard accretion models to the quasar spectra. They compare
the optical-UV slope predicted by the standard disk blackbody model to
the observed median slope of the quasar combined spectrum in LBQS
sample \cite{francis91}. Czerny et al \cite{czerny2003} considered
an universal shape of quasars and NLS1 spectra in comparison to the
simple disk-blackbody shape to illustrate the need for additional
processes to account for the observed flattening in the optical
spectrum with respect to the blackbody emission. They considered
irradiation and outflow as possible processes responsible for the
observed flattening of the slope (e.g. \cite{czerny2004b}). Dust may
also lead to observed flattening of the continuum
\cite{czerny2004,loska2004}.  Modeling a large sample of quasars in
the broad-band spectral domain is needed in order to constrain the
parameters of the accretion process.


\section{Summary}

We considered only the steady state accretion flow spectra and did not
really talk about the short and long term variability in theoretical
models and observations. Future global simulations should provide us
with an evolution of an accretion disk which is clearly
needed. Observations show AGN and quasar variability on different
timescales. Long term changes in the accretion state may affect our
understanding of black hole growth and evolution of structures in the
universe.

\begin{theacknowledgments}
The author would like to thank Bo{\.z}ena Czerny for discussion, Ma{\l
}gorzata Sobolewska for discussion and help in creating disk models
for the figures and Katherine Aldcroft for careful reading of the
manuscript. This research is funded in part by NASA contract
NAS8-39073 and through Chandra Award Number GO5-6113X, issued by the
Chandra X-Ray Observatory Center.

\end{theacknowledgments}


\bibliographystyle{aipproc}

\end{document}